\documentclass[a4paper,12pt]{article}
\usepackage{Stylefile}
\doublespace 

\title
{ %
\vspace*{3.0cm} \LARGE{\bf The Flow of Newtonian Fluids in Axisymmetric Corrugated Tubes} \vspace*{4.0cm} \\
}

\author{Taha Sochi\footnote{University College London, Department of Physics \& Astronomy, Gower Street, London, WC1E 6BT.
Email: t.sochi@ucl.ac.uk.} \vspace*{5.0cm}}

\begin{document}

\maketitle %
\pagenumbering{arabic}

\newpage
\phantomsection \addcontentsline{toc}{section}{Contents} %
\tableofcontents

\newpage
\phantomsection \addcontentsline{toc}{section}{List of Figures} %
\listoffigures

\newpage
\phantomsection \addcontentsline{toc}{section}{Abstract} \noindent
{\noindent \LARGE \bf Abstract} \vspace{0.5cm}\\
\noindent %

This article deals with the flow of Newtonian fluids through axially-symmetric corrugated tubes. An
analytical method to derive the relation between volumetric flow rate and pressure drop in laminar
flow regimes is presented and applied to a number of simple tube geometries of converging-diverging
nature. The method is general in terms of fluid and tube shape within the previous restrictions.
Moreover, it can be used as a basis for numerical integration where analytical relations cannot be
obtained due to mathematical difficulties.

\pagestyle{headings} %
\addtolength{\headheight}{+1.6pt}
\lhead[{Chapter \thechapter \thepage}]%
      {{\bfseries\rightmark}}
\rhead[{\bfseries\leftmark}]%
     {{\bfseries\thepage}} 
\headsep = 1.0cm               

\newpage
\section{Introduction} \label{Introduction}

Modeling the flow through corrugated tubes of various geometries is an important subject and has
many real-life applications. Moreover, it is required for modeling viscoelasticity, yield-stress
and the flow of Newtonian and non-Newtonian fluids through porous media \cite{SochiB2008,
SochiVE2009, SochiYield2010}. There are many previous attempts to model the flow through
capillaries of different geometries. However, they either apply to tubes of regular cross sections
\cite{Whitebook, SisavathJZ2001} or deal with very special cases. Most of these studies use
numerical mesh techniques such as finite difference and spectral methods to obtain numerical
results. Illuminating examples of these investigations are Kozicki \etal\ \cite{KozickiCT1966},
Miller \cite{Miller1972}, Oka \cite{Oka1973}, Williams and Javadpour \cite{WilliamsJ1980},
Phan-Thien \etal\ \cite{ThienGB1985, ThienK1987}, Lahbabi and Chang \cite{LahbabiC1986}, Burdette
\etal\ \cite{BurdetteCAB1989}, Pilitsis \etal\ \cite{PilitsisSB1991, PilitsisB1989}, James \etal\
\cite{JamesTKBP1990}, Talwar and Khomami \cite{TalwarK1992}, Koshiba \etal\ \cite{KoshibaMNS2000},
Masuleh and Phillips \cite{MasulehP2004}, and Davidson \etal\ \cite{DavidsonLC2008}.

\begin{figure}[!h]
\centering{}
\includegraphics
[scale=0.6] {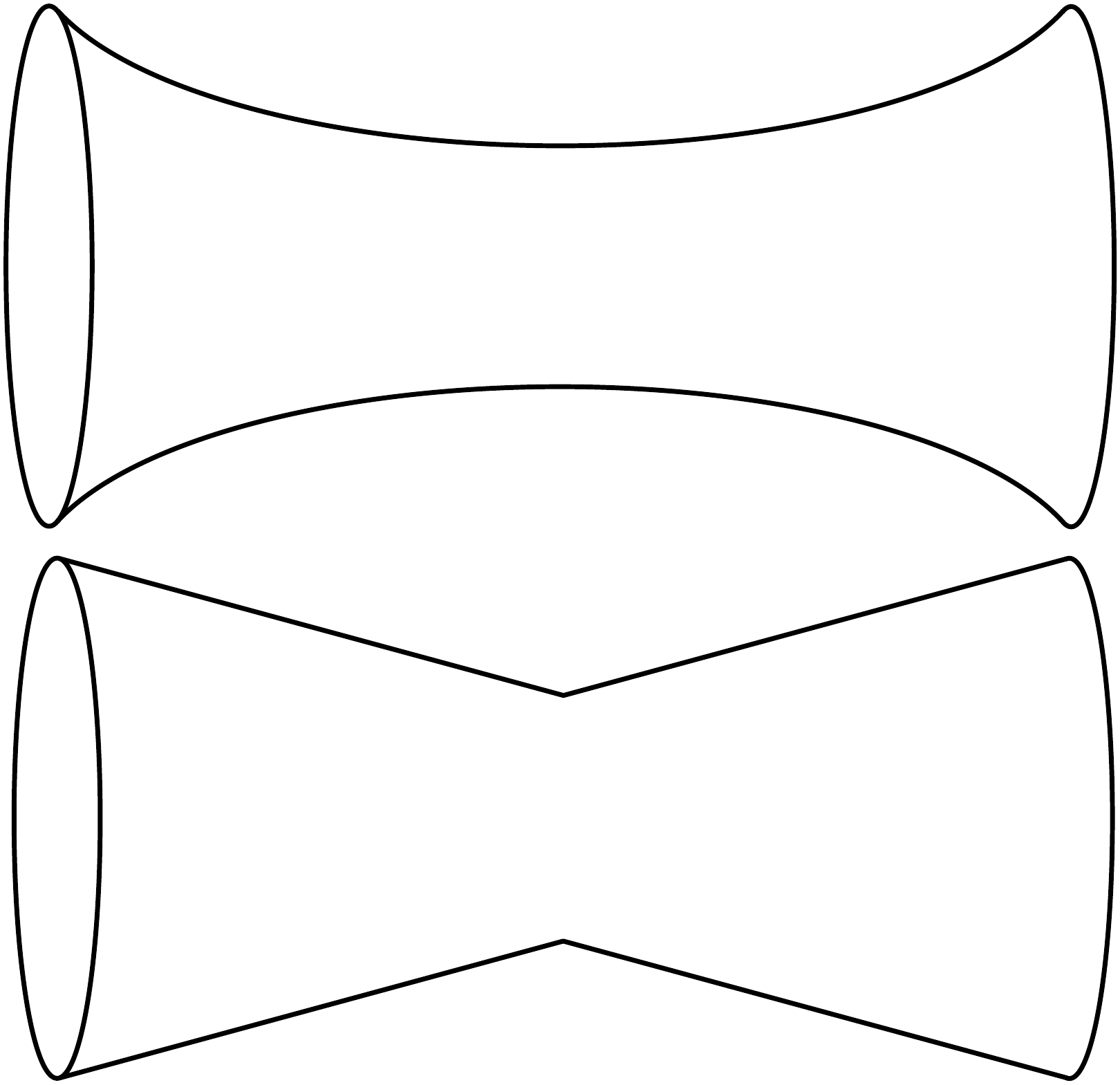} \caption{Profiles of converging-diverging axisymmetric capillaries.}
\label{Tubes}
\end{figure}

In the current paper we present an analytical method for deriving the relationship between pressure
drop and volumetric flow rate in corrugated tubes of circular but varying cross section, such as
those depicted schematically in Figure \ref{Tubes}. We also present several examples of the use of
this method to derive equations for Newtonian flow although the method is general and can be
applied to non-Newtonian flow as well. In the following derivations we assume a laminar flow of a
purely-viscous incompressible fluid where the tube corrugation is smooth and limited in magnitude
to avoid problematic flow phenomena such as vortices.

\section{Correlation Between Pressure Drop and Flow Rate} \label{}

A Newtonian fluid with a viscosity $\mu$ satisfy the following relation between stress $\tau$ and
strain rate $\dot{\gamma}$
\begin{equation}
\tau=\mu\dot{\gamma}\label{Newt}
\end{equation}

The Hagen-Poiseuille equation for a Newtonian fluid passing through a cylindrical pipe of constant
circular cross section, which can be derived directly from Equation \ref{Newt}, states that
\begin{equation}
Q=\frac{\pi r^{4}P}{8\mu x}\label{Pois}
\end{equation}
where $Q$ is the volumetric flow rate, $r$ is the tube radius, $P$ is the pressure drop across the
tube, $\mu$ is the fluid viscosity and $x$ is the tube length. A derivation of this relation can be
found, for example, in \cite{SochiB2008, Sochithesis2007}. On solving this equation for $P$, the
following expression for pressure drop as a function of flow rate is obtained

\begin{equation}
P=\frac{8Q\mu x}{\pi r^{4}}
\end{equation}

For an infinitesimal length, $\delta x$, of a capillary, the infinitesimal pressure drop for a
given flow rate $Q$ is given by

\begin{equation}
\delta P=\frac{8Q\mu\delta x}{\pi r^{4}}
\end{equation}

For an incompressible fluid, the volumetric flow rate across an arbitrary cross section of the
capillary is constant. Therefore, the total pressure drop across a capillary of length $L$ with
circular cross section of varying radius, $r(x)$, is given by

\begin{equation}
P=\frac{8Q\mu}{\pi}\int_{0}^{L}\frac{dx}{r^{4}}\label{PQEq}
\end{equation}

This relation will be used in the following sections to derive relations between pressure drop and
volumetric flow rate for a number of geometries of axisymmetric capillaries of varying cross
section with converging-diverging feature. The method can be equally applied to other geometries of
different nature.

\subsection{Conical Tube} \label{}

\begin{figure}[!h]
\centering{}
\includegraphics
[scale=0.8] {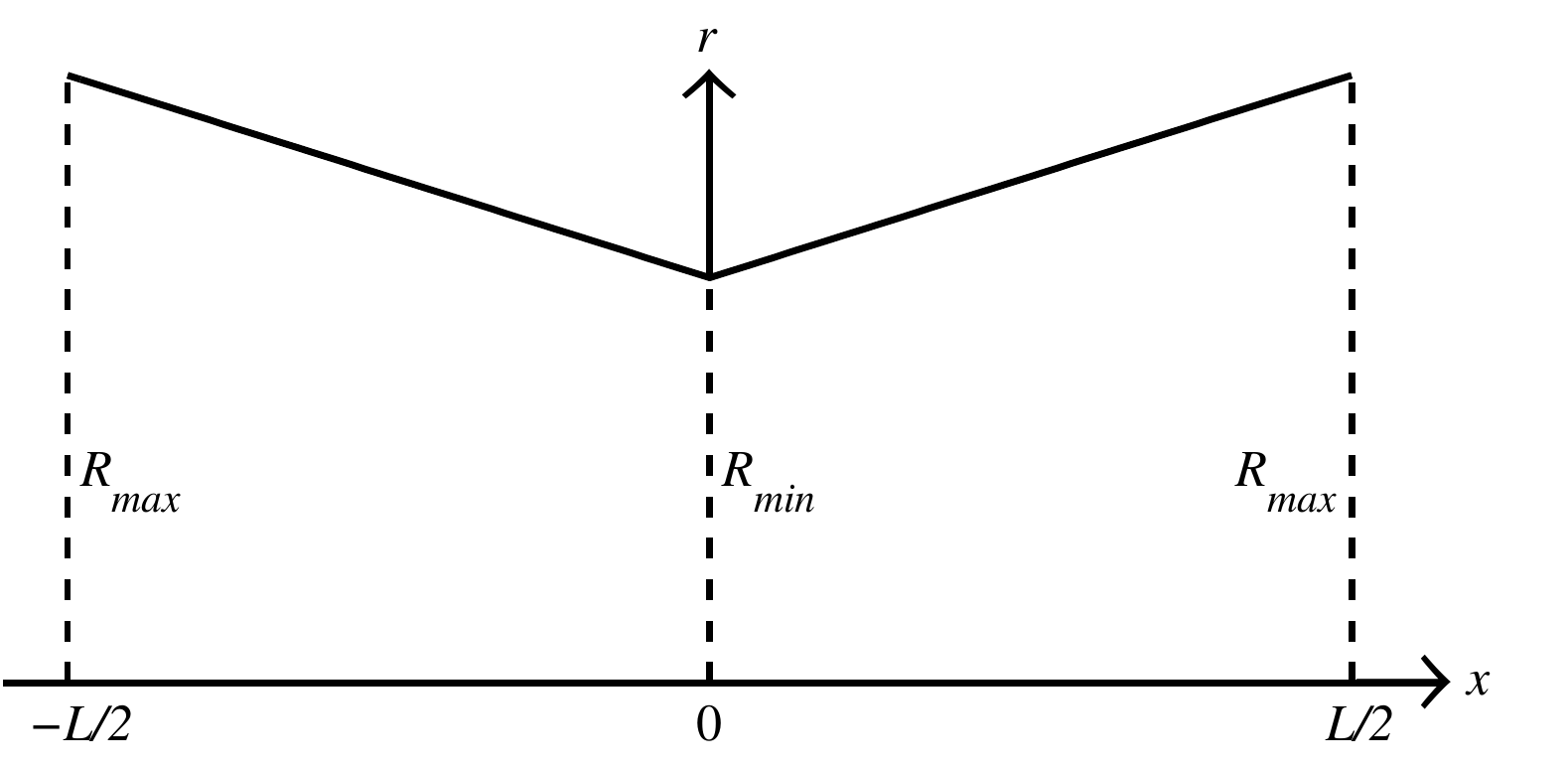} \caption{Schematic representation of the radius of a conically shaped
converging-diverging capillary as a function of the distance along the tube axis.} \label{Conic}
\end{figure}

For a corrugated tube of conical shape, depicted in Figure \ref{Conic}, the radius $r$ as a
function of the axial coordinate $x$ in the designated frame is given by

\begin{equation}
r(x)=a+b|x| \hspace{1cm} -L/2\le x \le L/2
\end{equation}
where

\begin{equation}
a=R_{min} \hspace{1cm} {\rm and} \hspace{1cm} b=\frac{2(R_{max}-R_{min})}{L}
\end{equation}

Hence, Equation \ref{PQEq} becomes

\begin{eqnarray}
  P&=&\frac{8Q\mu}{\pi}\int_{-L/2}^{L/2}\frac{dx}{(a+b|x|)^{4}} \\
   &=&\frac{8Q\mu}{\pi}\left[\frac{1}{3b(a-bx)^{3}}\right]_{-L/2}^{0}+\frac{8Q\mu}{\pi}\left[-\frac{1}{3b(a+bx)^{3}}\right]_{0}^{L/2} \\
   &=&\frac{16Q\mu}{\pi}\left[\frac{1}{3a^{3}b}-\frac{1}{3b(a+bL/2)^{3}}\right]
\end{eqnarray}

that is

\begin{equation} \label{fConic}
P=\frac{8LQ\mu}{3\pi(R_{max}-R_{min})}\left[\frac{1}{R_{min}^{3}}-\frac{1}{R_{max}^{3}}\right]
\end{equation}

\subsection{Parabolic Tube} \label{}

\begin{figure}[!h]
\centering{}
\includegraphics
[scale=0.8] {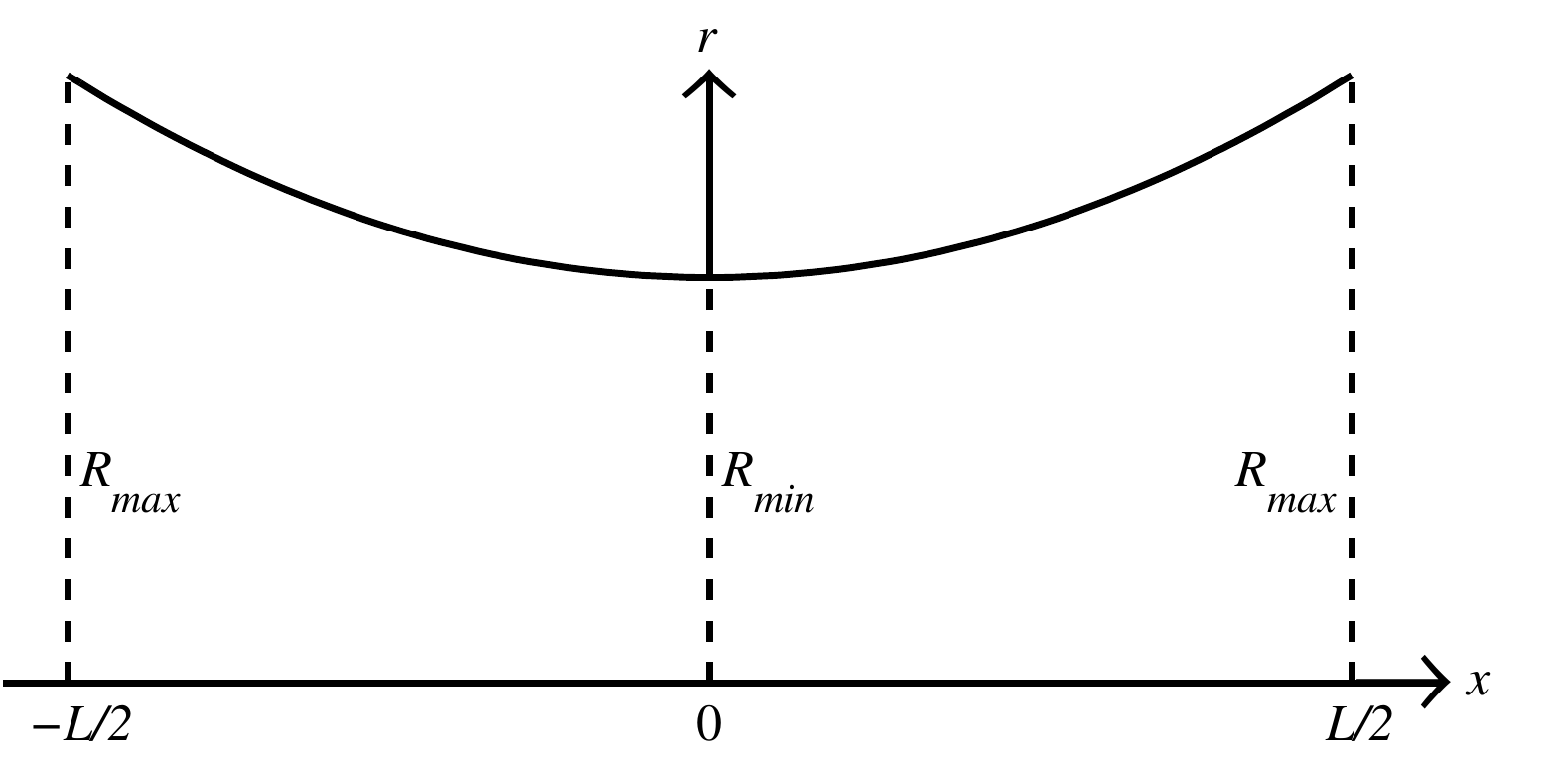} \caption{Schematic representation of the radius of a converging-diverging
capillary with a parabolic profile as a function of the distance along the tube axis.}
\label{Parabola}
\end{figure}

For a tube of parabolic profile, depicted in Figure \ref{Parabola}, the radius is given by

\begin{equation}
r(x)=a+bx^{2} \hspace{1cm} -L/2 \le x \le L/2
\end{equation}
where

\begin{equation}
a=R_{min} \hspace{1cm} {\rm and} \hspace{1cm} b=\left(\frac{2}{L}\right)^{2}(R_{max}-R_{min})
\end{equation}

Therefore, Equation \ref{PQEq} becomes

\begin{equation}
P=\frac{8Q\mu}{\pi}\int_{-L/2}^{L/2}\frac{dx}{\left(a+bx^{2}\right)^{4}}
\end{equation}

On performing this integration, the following relation is obtained

\begin{equation}
P=\frac{8Q\mu}{\pi}\left[\frac{x}{6a(a+bx^{2})^{3}}+\frac{5x}{24a^{2}(a+bx^{2})^{2}}+\frac{5x}{16a^{3}(a+bx^{2})}+\frac{5\arctan\left(x\sqrt{\frac{b}{a}}\right)}{16a^{7/2}\sqrt{b}}\right]_{-L/2}^{L/2}
\end{equation}

\begin{equation}
=\frac{8Q\mu}{\pi}\left[\frac{L}{6a[a+b(L/2)^{2}]^{3}}+\frac{5L}{24a^{2}[a+b(L/2)^{2}]^{2}}+\frac{5L}{16a^{3}[a+b(L/2)^{2}]}+\frac{10\arctan\left(\frac{L}{2}\sqrt{\frac{b}{a}}\right)}{16a^{7/2}\sqrt{b}}\right]
\end{equation}
that is

\begin{equation} \label{fParabolic}
P=\frac{4LQ\mu}{\pi}\left[\frac{1}{3R_{min}R_{max}^{3}}+\frac{5}{12R_{min}^{2}R_{max}^{2}}+\frac{5}{8R_{min}^{3}R_{max}}+\frac{5\arctan\left(\sqrt{\frac{R_{max}-R_{min}}{R_{min}}}\right)}{8R_{min}^{7/2}\sqrt{R_{max}-R_{min}}}\right]
\end{equation}

\subsection{Hyperbolic Tube} \label{}

For a tube of hyperbolic profile, similar to the profile in Figure \ref{Parabola}, the radius is
given by

\begin{equation}
r(x)=\sqrt{a+bx^{2}} \hspace{1cm} -L/2 \le x \le L/2 \hspace{1cm} a,b>0
\end{equation}
where

\begin{equation}
 a=R_{min}^{2} \hspace{1cm} {\rm and} \hspace{1cm} b=\left(\frac{2}{L}\right)^{2}(R_{max}^{2}-R_{min}^{2})
\end{equation}

Therefore, Equation \ref{PQEq} becomes

\begin{eqnarray}
P&=&\frac{8Q\mu}{\pi}\int_{-L/2}^{L/2}\frac{dx}{\left(a+bx^{2}\right)^{2}} \\
&=&\frac{8Q\mu}{\pi}\left[\frac{x}{2a(a+bx^{2})}+\frac{\arctan(x\sqrt{b/a})}{2a\sqrt{ab}}\right]_{-L/2}^{L/2}
\end{eqnarray}
that is

\begin{equation} \label{fHyperbolic}
P=\frac{4LQ\mu}{\pi}\left[\frac{1}{R_{min}^{2}R_{max}^{2}}+\frac{\arctan\left(\sqrt{\frac{R_{max}^{2}-R_{min}^{2}}{R_{min}^{2}}}\right)}{R_{min}^{3}\sqrt{R_{max}^{2}-R_{min}^{2}}}\right]
\end{equation}

\subsection{Hyperbolic Cosine Tube} \label{}

For a tube of hyperbolic cosine profile, similar to the profile in Figure \ref{Parabola}, the
radius is given by

\begin{equation}
r(x)=a\cosh(bx) \hspace{1cm} -L/2 \le x \le L/2
\end{equation}
where

\begin{equation}
 a=R_{min} \hspace{1cm} {\rm and} \hspace{1cm} b=\frac{2}{L}\arccosh\left(\frac{R_{max}}{R_{min}}\right)
\end{equation}

Hence, Equation \ref{PQEq} becomes

\begin{eqnarray}
P&=&\frac{8Q\mu}{\pi}\int_{-L/2}^{L/2}\frac{dx}{\left[a\cosh(bx)\right]^{4}} \\
&=&\frac{8Q\mu}{\pi}\left[\frac{\tanh(bx)\left[\mathrm{sech}^{2}(bx)+2\right]}{3a^{4}b}\right]_{-L/2}^{L/2}
\end{eqnarray}
that is

\begin{equation} \label{fCoshine}
 P=\frac{8LQ\mu}{3\pi R_{min}^{4}}\left[\frac{\tanh\left(\arccosh\left(\frac{R_{max}}{R_{min}}\right)\right)\left[\mathrm{sech}^{2}\left(\arccosh\left(\frac{R_{max}}{R_{min}}\right)\right)+2\right]}{\arccosh\left(\frac{R_{max}}{R_{min}}\right)}\right]
\end{equation}

\subsection{Sinusoidal Tube} \label{}

\begin{figure}[!h]
\centering{}
\includegraphics
[scale=0.8] {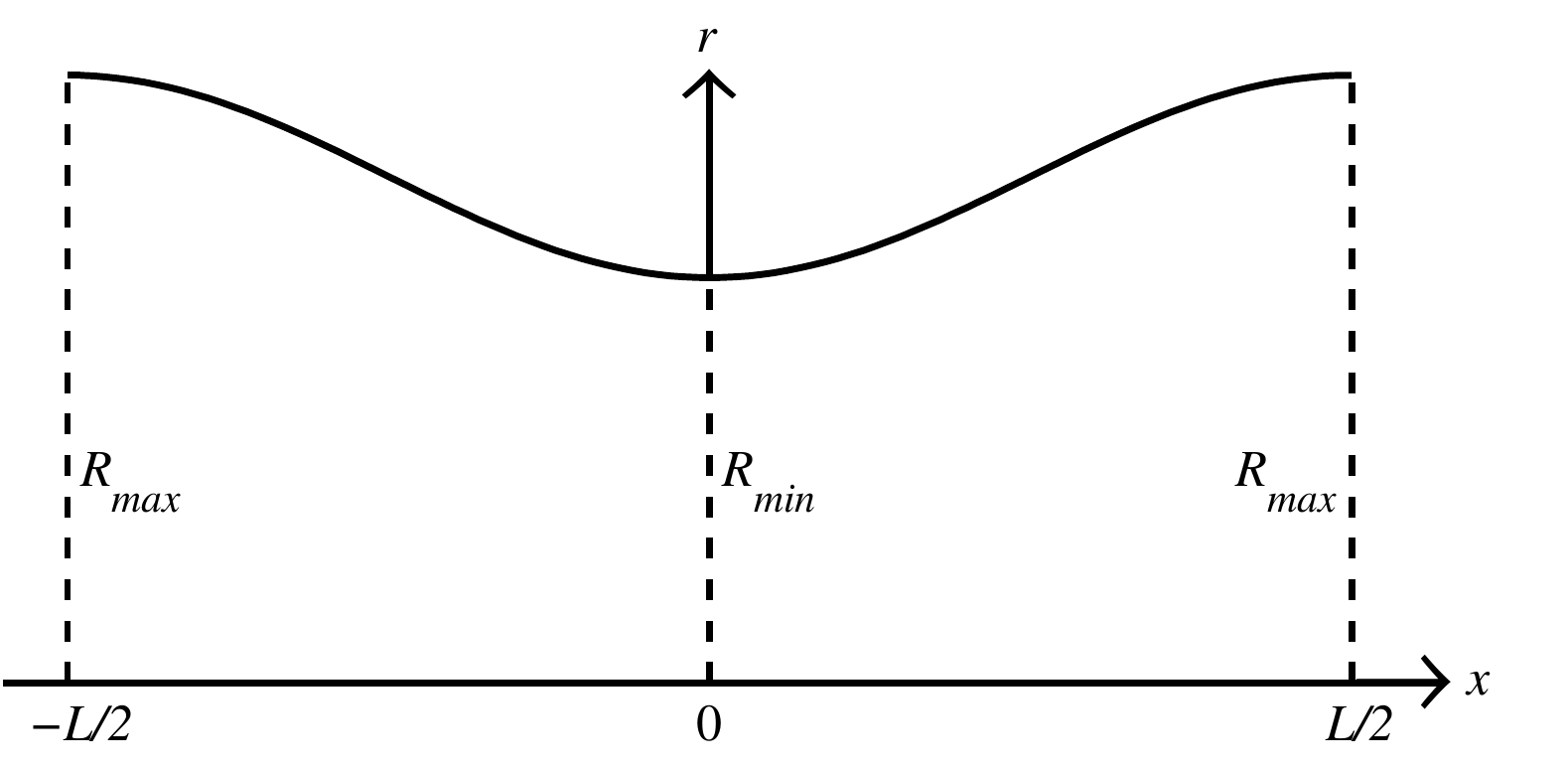} \caption{Schematic representation of the radius of a converging-diverging
capillary with a sinusoidal profile as a function of the distance along the tube axis.}
\label{Sinusoid}
\end{figure}

For a tube of sinusoidal profile, depicted in Figure \ref{Sinusoid}, where the tube length $L$
spans one complete wavelength, the radius is given by
\begin{equation}
r(x)=a-b\cos\left(kx\right) \hspace{1cm} -L/2 \le x \le L/2 \hspace{1cm} a>b>0
\end{equation}
where
\begin{equation}
a=\frac{R_{max}+R_{min}}{2} \hspace{1cm} b=\frac{R_{max}-R_{min}}{2} \hspace{1cm} \& \hspace{1cm}
k=\frac{2\pi}{L}
\end{equation}

Hence, Equation \ref{PQEq} becomes

\begin{equation}
P=\frac{8Q\mu}{\pi}\int_{-L/2}^{L/2}\frac{dx}{\left[a-b\cos(kx)\right]^{4}}
\end{equation}

On performing this integration, the following relation is obtained
\begin{equation}
P=\frac{8Q\mu}{\pi b^{4}k}\left[I\right]_{-L/2}^{L/2}
\end{equation}
where
\begin{eqnarray}
\nonumber
I&=&\frac{(6A^{3}+9A)}{3(A^{2}-1)^{7/2}}\arctan\left(\frac{(A-1)\tan(\frac{kx}{2})}{\sqrt{A^{2}-1}}\right)-\frac{(11A^{2}+4)\sin(kx)}{6(A^{2}-1)^{3}[A+\cos(kx)]} \\
 & & -\frac{5A\sin(kx)}{6(A^{2}-1)^{2}[A+\cos(kx)]^{2}}-\frac{\sin(kx)}{3(A^{2}-1)[A+\cos(kx)]^{3}}
\end{eqnarray}
\begin{equation}
\& \hspace{1cm} A=\frac{R_{max}+R_{min}}{R_{min}-R_{max}}
\end{equation}

On taking $\lim_{x\rightarrow-\frac{L}{2}^{+}}I$ and $\lim_{x\rightarrow\frac{L}{2}^{-}}I$ the
following expression is obtained
\begin{eqnarray}
P&=&\frac{8Q\mu}{\pi b^{4}k}\left[-\frac{(6A^{3}+9A)}{3(A^{2}-1)^{7/2}}\frac{\pi}{2}-\frac{(6A^{3}+9A)}{3(A^{2}-1)^{7/2}}\frac{\pi}{2}\right] \\
 &=&-\frac{8Q\mu(6A^{3}+9A)}{3b^{4}k(A^{2}-1)^{7/2}}
\end{eqnarray}

Since $A<-1$, $P>0$ as it should be. On substituting for $A$, $b$ and $k$ in the last expression we
obtain
\begin{equation}
P=\frac{LQ\mu(R_{max}-R_{min})^{3}\left[2\left(\frac{R_{max}+R_{min}}{R_{max}-R_{min}}\right)^{3}+3\left(\frac{R_{max}+R_{min}}{R_{max}-R_{min}}\right)\right]}{2\pi(R_{max}R_{min})^{7/2}}
\end{equation}
that is
\begin{equation} \label{fSinusoid}
P=\frac{LQ\mu\left[2(R_{max}+R_{min})^{3}+3(R_{max}+R_{min})(R_{max}-R_{min})^{2}\right]}{2\pi(R_{max}R_{min})^{7/2}}
\end{equation}

\vspace{0.5cm}

It is noteworthy that all these relations (i.e. Equations \ref{fConic}, \ref{fParabolic},
\ref{fHyperbolic}, \ref{fCoshine} and \ref{fSinusoid}), are dimensionally consistent. Moreover,
they have been extensively tested and verified by numerical integration.

\newpage

\section{Conclusions} \label{Conclusions}

The current paper proposes an analytical method for deriving mathematical relations between
pressure drop and volumetric flow rate in axially symmetric corrugated tubes of varying cross
section. This method is applied on a number of converging-diverging capillary geometries in the
context of Newtonian flow. The method can be equally applied to some cases of non-Newtonian flow
within the given restrictions. It can also be extended to include other regular but
non-axially-symmetric geometries. The method can be used as a basis for numerical integration when
analytical solutions are out of reach due to mathematical complexities.

\vspace{1cm}

\phantomsection \addcontentsline{toc}{section}{Nomenclature} %
{\noindent \LARGE \bf Nomenclature} \vspace{0.5cm}

\begin{supertabular}{ll}
$\sR$                 & strain rate (s$^{-1}$) \\
$\mu$                 & fluid viscosity (Pa.s) \\
$\sS$                 & stress (Pa) \\
\\
$L$                   & tube length (m) \\
$P$                   & pressure drop (Pa) \\
$Q$                   & volumetric flow rate (m$^{3}$.s$^{-1}$) \\
$r$                   & tube radius (m) \\
$R_{max}$             & maximum radius of corrugated tube (m) \\
$R_{min}$             & minimum radius of corrugated tube (m) \\
$x$                   & axial coordinate (m) \\
\end{supertabular}

\newpage
\phantomsection \addcontentsline{toc}{section}{References} %
\bibliographystyle{unsrt}
\bibliography{Biblio}

\end{document}